\documentclass[twocolumn,amsmath,amssymb,prx,showpacs]{revtex4}
\usepackage{graphicx}
\usepackage{times}
\usepackage{pstricks}

\begin{document}

\title{Network topology: detecting topological phase transitions in the Kitaev chain and the rotor plane}

\author{Chung-Pin Chou$^{1}$}\email{cpc63078@gmail.com}
\author{Ming-Chiang Chung$^{2,3}$}\email{mingchiangha@phys.nchu.edu.tw}

\affiliation{$^{1}$Beijing Computational Science Research Center,
Beijing 100084, China}\affiliation{$^{2}$Department of Physics,
National Chung Hsing University, Taichung 40227,
Taiwan}\affiliation{$^3$Physics Division, National Center for
Theoretical Science, Hsinchu, 30013, Taiwan}

\begin{abstract}
We propose a novel network measure of topological invariants, called
small-worldness, for identifying topological phase transitions of
quantum and classical spin models. Small-worldness is usually
defined in the study of social networks based on the best known
discovery that one can find a short chain of acquaintances
connecting almost any two people on the planet. Here we demonstrate
that the small-world effect provides a useful description to
distinguish topologically trivial and non-trivial phases in the
Kitaev chain and accurately capture the Kosterlitz-Thouless
transition in the rotor plane. Our results further suggest that the
small-worldness containing both locality and non-locality of the
network topology can be a practical approach to extract
characteristic quantities of topological states of matter.
\end{abstract}


\maketitle

\section{Introduction}\label{intro}
Determining the topological order of an interacting quantum system
from its microscopic many-body entanglement is one of the recent
goals of condensed matter theory. Traditional phases of matter and
phase transitions are usually distinguished by local order
parameters. Consider, for instance, a second-order phase transition,
the critical point is accompanied by a diverging correlation length
in Landau's symmetry breaking framework \cite{SachdevBook99}.
However, it becomes clear that some topologically ordered phases do
not fall into this framework, such as fractional quantum Hall states
\cite{LaughlinPRL83}, quantum spin liquids \cite{AndersonMRB73} and
recently discovered topological insulators
\cite{HasanRMP10,QiRMP11}. Moreover, characterizing topological
phase transitions between them is a difficult task due to the
absence of local order parameters \cite{XGWenIJMP90,CarrBook11}.
There are recent advances in diagnosing the presence of topological
order from knowledge of many-body ground states, according to
entanglement entropy \cite{KitaevPRL06,LevinPRL06}, entanglement
spectrum \cite{LiPRL08,QiPRL12}, modular transformation
\cite{ZhangPRB12}, Chern number \cite{ThoulessPRL82}, $Z_{2}$
topological invariant \cite{KitaevPU01,KanePRL05} and momentum
polarization \cite{TuPRB13}. They all have brought us closer to
being able to tackle such important questions.

Another interesting question is the nature of phase transitions
between topologically ordered states. Some of those methods
mentioned above to identify a topological phase are usually not
unique when several topological phases can have the same topological
invariants \cite{WenCMP13}. Furthermore, these methods cannot
provide more complete information about all phase transitions
incorporating traditional Landau's symmetry breaking picture. To
address this issue, we attempt to map a physical model to a network
possibly offers an alternative understanding of the essence of phase
transitions in condensed matter physics.

The complex network theory originating from the graph theory in
mathematics has become one standard tool to analyze the structure
and dynamics of real-world systems, which consist of overwhelming
information
\cite{AlbertRMP02,DorogovtsevAIP02,NewmanSIAM03,BoccalettiPR06,DorogovtsevRMP08}.
The building blocks of the complex network include nodes (system
elements) and links (the relation between two elements). Via the
unique patterns of connections, the essential features of a
collection of interacting elements can be unveiled by the direct
visualization and the topological analysis. Its application is
prevailing in many fields, such as sociology, biology, informatics
and many other interdisciplinary studies \cite{SethnaBook06}. Thus,
this generic description is reasonably applicable to condensed
matter systems.

A naive question in the condensed matter theory and the application
of network analysis is whether or not the topological phase
transitions can be detected by complex network topology. Out of many
properties in real-world networks, the small-world effect is a
common phenomenon, which is characterized by a small average length
of the shortest paths between two nodes. The characteristics of
strong clustering and shortest path length, proposed by Watts and
Strogatz decades ago \cite{WattsNat98}, establishes enough
long-range connections in network space. In light of extensive
studies in various real-world networks
\cite{AlbertNat99,NewmanPNAS01,DoddsSci03,BackstromProc12}, it will
be interesting to explore the point of view of the small-world
effect in many-body systems.

The focus of the present paper is to construct the weighted networks
for the Kitaev chain and the two-dimensional (2D) classical rotor
model. The weights of network links we define carry quantum and
classical correlations between lattice sites in the Kitaev chain and
the rotor plane, respectively. An interesting observation is that
the two topologically distinct phases in the Kitaev chain can be
distinguished by the novel network topology instead of typical
topological invariants. The topology arising from the weighted
networks is illustrated by the small-world phenomenon. Using the
network property in the rotor plane, we find that it can also
extract the Kosterlitz-Thouless (KT) transition point obtained by
conventional quantities, {\it e.g.} spin stiffness. In order to
further quantify the small-world network, we propose a network
quantity, small-worldness, as an order parameter in these two
models. The small-worldness exhibits an obvious change while these
systems undergo a topological phase transition. These phases of
matter are encoded in the network representation, from which we can
extract the universal properties of the topological phase
transitions.

The remainder of this paper is organized as follows. In
Sec.\ref{netp}, we introduce several common network measures
computed in this paper. Following this, in Sec.\ref{1dq}, we display
the network topology and the small-world effect in the Kitaev chain.
In Sec.\ref{2dc}, we discuss the similar results obtained in the
classical rotor plane. Finally, we provide a conclusion and an
outlook in Sec.\ref{conc}.

\section{Network measures}\label{netp}
In the language of complex network, each network of $N$ nodes is
described by its $N\times N$ adjacency matrix representation
$\hat{A}$ \cite{AlbertRMP02}. A number of real systems, {\it e.g.}
transportation networks, neural networks and so on, are better
captured by the weighted network in which the links use weights to
quantify their strengths. In these two models, we consider lattice
sites as nodes of the weighted network, with each weighted link
between nodes $i$ and $j$ expressed by the element of the adjacency
matrix $\hat{A}_{ij}$. The link carries the weight containing
detailed information about the relationship between particles at
different sites. Thus the weights of network links are assigned by
quantum and classical correlations between particles in the Kitaev
chain and the rotor plane, respectively. Note that the symmetric
adjacency matrix $\hat{A}_{ij}$ in our examples only has real
entries. Following the convention in weighted complex networks
\cite{SaramakiPRE07}, here we take the absolute value of
$\hat{A}_{ij}$ normalized by the maximum weight in the network,
$\frac{|\hat{A}_{ij}|}{\max{|\hat{A}_{ij}|}}$.

Two concepts, the clustering coefficient and the path length, play a
key role in the development of network science in the last decades.
They can be easily evaluated from the adjacency matrix
$\hat{A}_{ij}$. The former refers to the local property of a
network. More precisely, if the neighbors of a given node connect to
each other, a local cluster will be formed in an unweighted network.
As for a weighted network, the degree of clustering of the network
can be captured by the weighted clustering coefficient $C$
\cite{OnnelaPRE05}, given by
\begin{eqnarray}
C=\frac{1}{N}\sum_{i}\frac{\left[\left(\hat{A}^{(1/3)}\right)^{3}\right]_{ii}}{k_{i}(k_{i}-1)}.\label{e:equ3}
\end{eqnarray}
Here $\hat{A}^{(n)}$ is a matrix obtained from $\hat{A}$ by taking
the n-th root of its individual elements.
$k_{i}(=\sum_{j=1}^{N}\hat{A}_{ij})$ represents the node strength
for each node $i$. Notice that according to the definition of the
adjacency matrix, the unweighted and weighted clustering
coefficients are equivalent as the weights become binary. Thus one
can regard Eq.(\ref{e:equ3}) as the probability with two neighbors
of a randomly selected node linking to each other.

We now turn to another fundamental concept in graph theory which is
the shortest path of nodes. Physical distance in a network is
usually irrelevant, and should be replaced by the path length. A
path is a route that runs along the links of a network. The path
length is thus defined by the inverse of the link weights the path
contains. This definition seizes the intuitive idea that strongly
coupled nodes are close to each other. It is noteworthy, then, that
the path length can represent the non-local property of a network.
Hence the path length $D$ is the average of the shortest path
lengths in a network \cite{NewmanPRB01}, defined as
\begin{eqnarray}
D=\frac{2}{N(N-1)}\sum_{i<j}\hat{d}_{ij},\label{e:equ4}
\end{eqnarray}
where $\hat{d}_{ij}$ is the sum of $\hat{A}_{\mu\nu}^{-1}$ along the
shortest weighted path including nodes $\mu$ and $\nu$ between nodes
$i$ and $j$. Here we compute the matrix $\hat{d}_{ij}$ by using the
well-known graph search algorithm which is called Dijkstra's
algorithm \cite{CormenBook09}.

In most real networks, a majority of nodes may not be neighbors but
can reach each other by a small number of steps which means
relatively small path length. This is called small-world phenomenon
\cite{WattsBook99}. However, although the random network also shows
the small-world effect, it still fails to reproduce some important
features of real networks, such as clustering. A small-world network
including not only high clustering but also short path length has
thus been introduced to describe many real networks by Watts and
Strogatz \cite{WattsNat98}. A network measure of the small-world
property called "small-worldness" has been proposed as well
\cite{HumphriesPLos08}. The definition is based on the maximal
tradeoff between high clustering (large $C$) and short path length
(small $D$). We can further define the small-worldness as
\begin{eqnarray}
S\equiv\frac{C}{D}.\label{e:equ5}
\end{eqnarray}
A network with larger $S$ has a higher small-world level
\cite{smallworld}. If a network is complete, {\it i.e.} all nodes
are connected with equal link weights, both $C$ and $D$ will
approach $1$, and then $S\rightarrow1$ (which means an extremely
small world). Therefore, the small-worldness can simultaneously
contain the local and non-local properties of a given network
topology. Later we will show that the small-worldness behaves as an
order parameter in the topological phase transitions.

\section{Kitaev chain}\label{1dq}
We start off with one-dimensional (1D) quantum Ising model in a
chain of length $L$ with periodic boundaries \cite{SachdevBook99}.
The Ising Hamiltonian with transverse magnetic field is written as
\begin{eqnarray}
H_{Ising}=-g\sum_{i}\hat{\sigma}_{i}^{z}-\sum_{i}\hat{\sigma}_{i}^{x}\cdot\hat{\sigma}_{i+1}^{x},\label{e:equ6}
\end{eqnarray}
where $\hat{\sigma}_{i}^{x}$ and $\hat{\sigma}_{i}^{z}$ are the
usual Pauli matrices and $g$ represents a dimensionless magnetic
field. In the Ising model, the $Z_{2}$ spin reflection symmetry is
spontaneously broken while the quantum phase transition takes place
at the critical field ($g_{c}=1$). For much larger magnetic field
$g$, the ground state is a quantum paramagnet with all spins
polarized along the field, whereas for small $g$, there are two
degenerate ferromagnetic ground states with all spins pointing
either "up" or "down" perpendicular to the magnetic field.

The 1D quantum Ising model can be re-written in terms of spinless
fermion by using the Jordan-Wigner transformation
\cite{SachdevBook99,PfeutyAP70}. Through the transformation, the
model turns into the Kitaev chain, a 1D $p$-wave superconductor with
the nearest-neighbor pairing term equal to the hopping term,
described by the Hamiltonian
\begin{eqnarray}
H_{Kitaev}=-\sum_{i}\left(\hat{c}_{i}^{\dag}\hat{c}_{i+1}+\hat{c}_{i}^{\dag}\hat{c}_{i+1}^{\dag}+H.c.\right)+\mu\hat{c}_{i}^{\dag}\hat{c}_{i},\label{e:equ6-1}
\end{eqnarray}
where $\mu$($\equiv2g$) is chemical potential. Hereafter the
particle-hole symmetry allows us to consider only the case
$\mu\geq0$. The simplest superconducting (SC) model system shows the
two-fold ground-state degeneracy stemming from an unpaired Majorana
fermion at the end of the chain with open boundary conditions (OBC).
A characteristic feature of the topological order is thus encoded in
the Majorana zero mode \cite{AliceaRPP12}. Kitaev showed that this
model has two phases sharing the same physical symmetries: a
topologically trivial phase for $\mu>2$ and a topologically
non-trivial phase for $\mu<2$ \cite{KitaevPU01}. The transition
between them is the topological phase transition identified by the
presence or absence of unpaired Majorana fermions localized at each
end.

The Hamiltonian in momentum space is quadratic of fermionic
operators $\hat{c}_{\mathbf{k}}$, which has the form:
\begin{eqnarray}
\sum_{\mathbf{k}}\left(
 \begin{array}{cc}
  \hat{c}_{\mathbf{k}}^{\dag} & \hat{c}_{-\mathbf{k}} \\
  \end{array}
  \right)\left(
         \begin{array}{cc}
         -\frac{\mu}{2}-\cos{\mathbf{k}} & -i\sin{\mathbf{k}} \\
         i\sin{\mathbf{k}} & \frac{\mu}{2}+\cos{\mathbf{k}} \\
         \end{array}
        \right)\left(
         \begin{array}{c}
         \hat{c}_{\mathbf{k}} \\
         \hat{c}_{-\mathbf{k}}^{\dag} \\
         \end{array}\right).\label{e:equ7}
\end{eqnarray}
Note that the periodic boundaries of the spin chain become
anti-periodic boundary condition for the spinless fermion. By using
the standard Bogoliubov transformation,
$\gamma_{\mathbf{k}}=\cos{(\theta_{\mathbf{k}}/2)}\hat{c}_{\mathbf{k}}-i\sin{(\theta_{\mathbf{k}}/2)}\hat{c}_{-\mathbf{k}}^{\dag}$
where
$\tan{\theta_{\mathbf{k}}}=-\sin{\mathbf{k}}/(\frac{\mu}{2}+\cos{\mathbf{k}}
)$, Eq.({\ref{e:equ7}}) can be diagonalized. The excitation spectrum
of the form,
$E_{\mathbf{k}}=\sqrt{\left(2\cos{\mathbf{k}}+\mu\right)^{2}+\sin^{2}{\mathbf{k}}}$,
remains fully gapped except for the critical point $\mu_{c}$($=2$).
The SC ground state is the state annihilated by all
$\gamma_{\mathbf{k}}$ \cite{ChungPRB00}, given by
\begin{eqnarray}
|\Psi_{GS}\rangle=e^{\frac{1}{2}\sum_{i,j}G_{ij}\hat{c}_{i}^{\dag}\hat{c}_{j}^{\dag}}|0\rangle.\label{e:equ8}
\end{eqnarray}
$G_{ij}$ represents the pairing amplitude defined by the Fourier
transform of $\tan{(\theta_{\mathbf{k}}/2)}$. It has been proven
that for the Kitaev chain the reduced density matrices can be
determined from the properties of the pairing amplitude
\cite{PeschelJPA03}. Thus the link weights of the Kitaev-chain
network are assigned by the pairing amplitude in which the non-local
property between spinless fermions is concealed.

\begin{figure}
\center
\includegraphics[height=2.4in,width=3.2in]{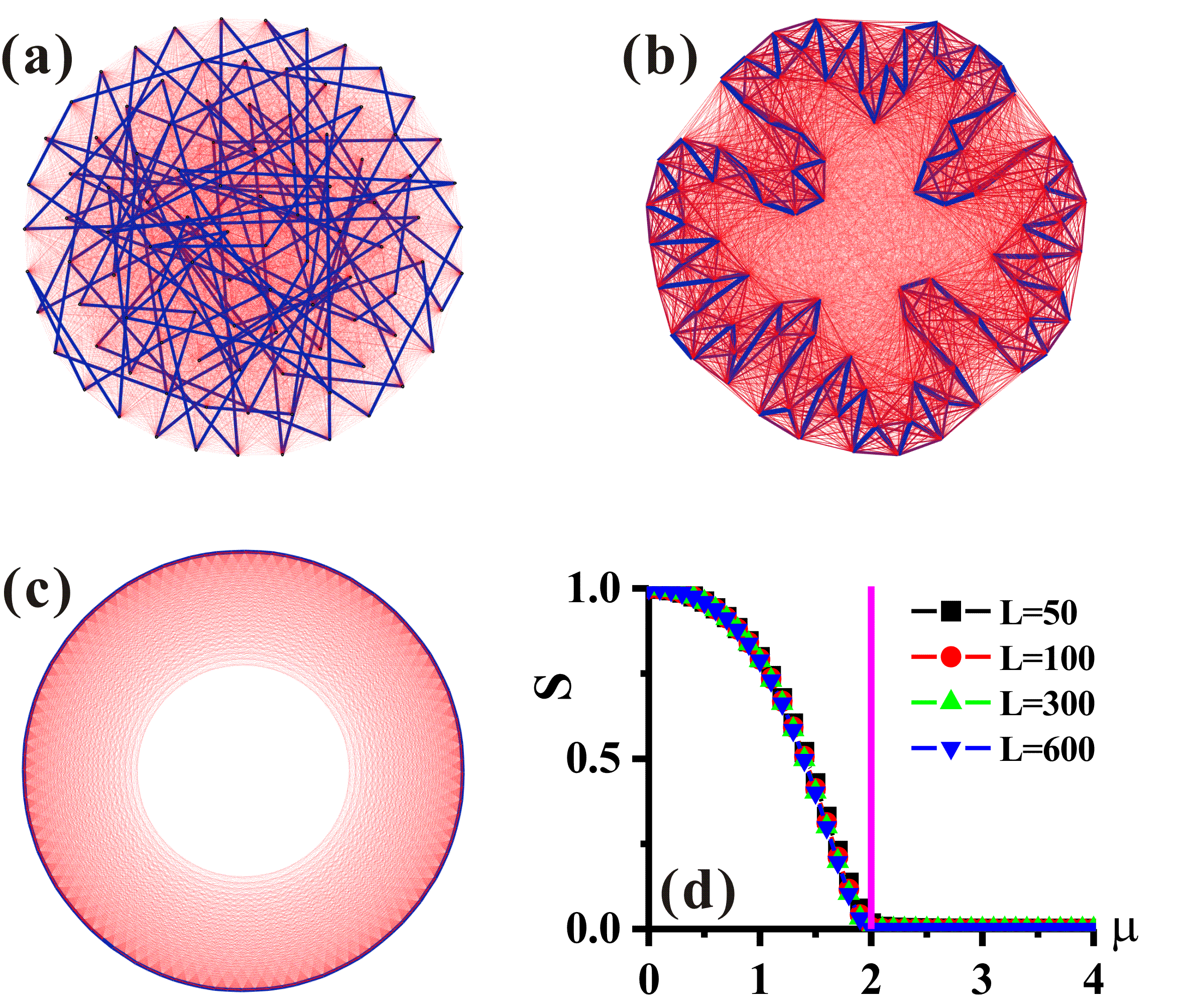}
\caption{Network representations of the Kitaev chain for different
chemical potential: (a) $\mu=0.2$, (b) $\mu=2$ and (c) $\mu=4$. The
chain length $L=100$. The thickness of links represents the
amplitude of $G_{ij}$. Color scale: Blue (Red) indicates the largest
(smallest) link weights. The network graphs are generated using the
force-directed algorithm \cite{gephi}. (d) Small-worldness $S$ of
the Kitaev chain as a function of chemical potential $\mu$ for
different chain length $L$. The pink vertical line indicates the
critical point $\mu_{c}$.}\label{fig1}
\end{figure}

The Kitaev chain admits gapless excitations only when the Fermi
level coincides with the top of the conduction band ($\mu=\mu_{c}$).
The two gapped phases are intuitively different in the regimes with
$\mu<\mu_{c}$ and $\mu>\mu_{c}$ owing to closing the excitation gap
at the critical point. Figure \ref{fig1}(a)-(c) show network
topologies at different chemical potential $\mu$ in the Kitaev
chain. Below the critical point $\mu_{c}$, the SC ground state
corresponds to a weak pairing regime in which the size of the Cooper
pair is infinite. For $\mu=0$, which gives rise to the unpaired
Majorana fermions at each end of the chain with OBC, the
recombination of the unpaired Majorana fermions generates an
ordinary fermion with a highly non-local property. We here observe a
trivial complete network at $\mu=0$, where each node is connected to
all other nodes with equal link weights. As increasing $\mu$, the
Majorana end states decay exponentially into the bulk of the chain
and the Cooper pair size seems to be slowly reduced. The
topologically non-trivial phase thus displays irregular patterns of
the network shown in Fig.\ref{fig1}(a).

At the critical point, the critical phase has power-law correlations
at large distances. Interestingly, we find that the nodes with the
largest link weight begin to form a "chain-like" structure in
Fig.\ref{fig1}(b). The obvious change of the topology of the network
is intimately related to the critical behavior observed in real
space. Above the critical point, {\it i.e.} $\mu>\mu_{c}$, the
ground state is instead in a strong pairing regime where the pairing
amplitude is exponentially decaying with distances. The Cooper pairs
form molecules from two fermions bound in real space over a length
scale. Exponentially decaying pairing amplitude in real space
results in the strongest links between neighboring nodes in network
space. In Fig.\ref{fig1}(c), the topologically trivial phase
demonstrates that a clear ring structure comprised of the nodes with
the largest link weight emerges in the network pond. Similar physics
would appear in the well-known "BEC-BCS crossover" in $s$-wave
superconductors without any sharp transition \cite{ChenPR05,CPC14}.

As mentioned above, the weak and strong pairing phases are obviously
distinct and separated by a topological phase transition at which
the bulk gap closes. One can express the topological invariant
distinguishing them, such as the Majorana number \cite{KitaevPU01}.
However, we here propose an entirely different point of view from
complex network analysis to detect the topological phase transition.
We calculate the small-worldness $S$ in the Kitaev chain.
Surprisingly, in Fig.\ref{fig1}(d) the small-worldness drops to zero
when the SC system comes from the topologically non-trivial phase to
the topologically trivial phase across the critical point $\mu_{c}$.
In particular, it shows less finite size dependence than other
network measures, e.g. the weighted clustering coefficient $C$ (not
shown). As a result, we illustrate that the network topology enables
the small-worldness, akin to an order parameter in the theory of
conventional phase transitions, to expose the change of nontrivial
topology inherent in the weak pairing regime.

\begin{figure}
\center
\includegraphics[height=3in,width=2.6in]{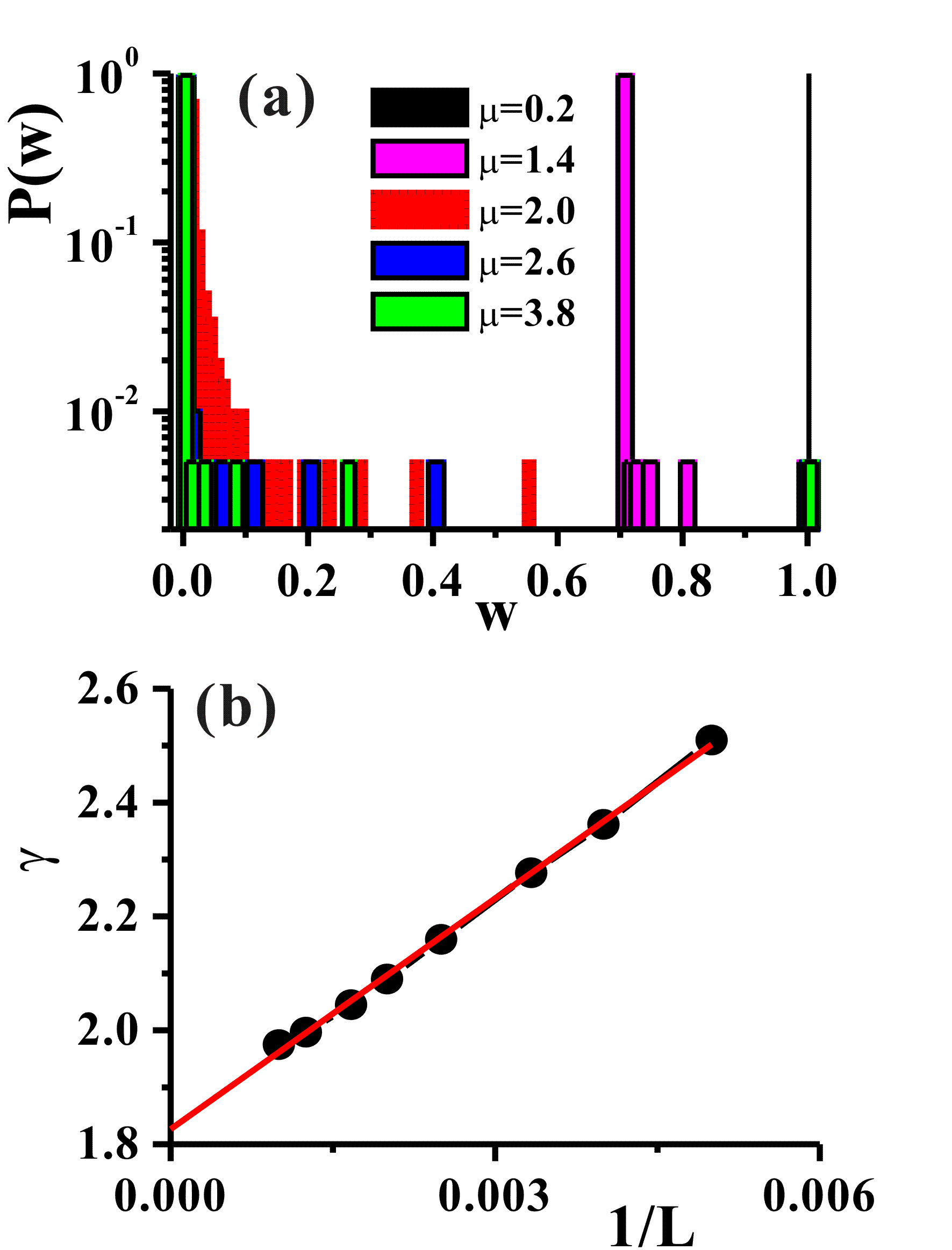}
\caption{(a) The probability distribution $P(w)$ of the link weight
$w$ for different $\mu$. The bin size is chosen as $0.01$ for clear
demonstrations. The chain length $L=400$. (b) The finite size
scaling of the critical exponent $\gamma$ determined from the
fitting formula $S=(\mu_{c}-\mu)^{\gamma}$ near the critical point
$\mu_{c}^{-}$.}\label{fig2}
\end{figure}

In order to further explore the implications of the network
topology, we investigate the probability distribution $P(w)$ of the
weights of network links in the Kitaev chain (see
Fig.\ref{fig2}(a)). In the case of $\mu<\mu_{c}$, the link weights
$w$ distribute like a delta function due to the long-range Cooper
pairs in real space. Namely, the link weights homogeneously
distribute in the network space. As further increasing $\mu$, the
position of the peak of the distribution are moved left but still
remains nearly homogeneous. It is noteworthy that the distribution
at the critical point possesses a decaying function with a heavy
tail and much broader width than others at different $\mu$. Hence
the weight distribution of network links becomes more heterogeneous.
On the other hand, in the strong pairing regime ($\mu>\mu_{c}$) the
weight distribution of network links moves to the weight $w\sim0$
and recovers the sharp peak. The tail of the distribution now looks
much more heterogeneous as a result of the formation of
"molecule-like" Cooper pairs. The sudden change of the weight
distribution at the critical point makes it easier to classify the
network links in both topologically trivial and non-trivial phases
so that we can have a clear order parameter to identify the phase
transition.

To examine the novel idea at the critical point, we compare the
small-worldness with the common order parameter in the 1D quantum
Ising model, spontaneous magnetization $M$, defined as
\begin{eqnarray}
M=\left\langle\left|\sum_{i}\hat{\sigma}_{i}^{z}\right|\right\rangle.\label{e:eq1}
\end{eqnarray}
It is well-known that the quantum phase transition in the chain
belongs to the universality class of the 2D classical Ising model,
which has been analytically solved by Onsager \cite{OnsagerPR44}. In
the thermodynamic limit, the singular behaviors of the spontaneous
magnetization near the critical external magnetic field $g_{c}$ can
be described by the scaling form: $(g_{c}-g)^{\beta}$ with the
critical exponent $\beta=\frac{1}{8}$. In Fig.\ref{fig2}(b), we
extrapolate the critical exponent $\gamma(\simeq1.83)$ of the
small-worldness $S\propto(\mu_{c}-\mu)^{\gamma}$ from the finite
size analysis, which is unexpectedly bigger than $\beta$. Notably,
$\gamma$ is very close to the critical exponent describing the
divergence of magnetic susceptibility, whose value is $\frac{7}{4}$.
This result strongly suggests that near the critical point the
small-worldness behaves as the second derivative of the free energy
with respect to some twist.

\begin{figure}
\center
\includegraphics[height=3in,width=2.6in]{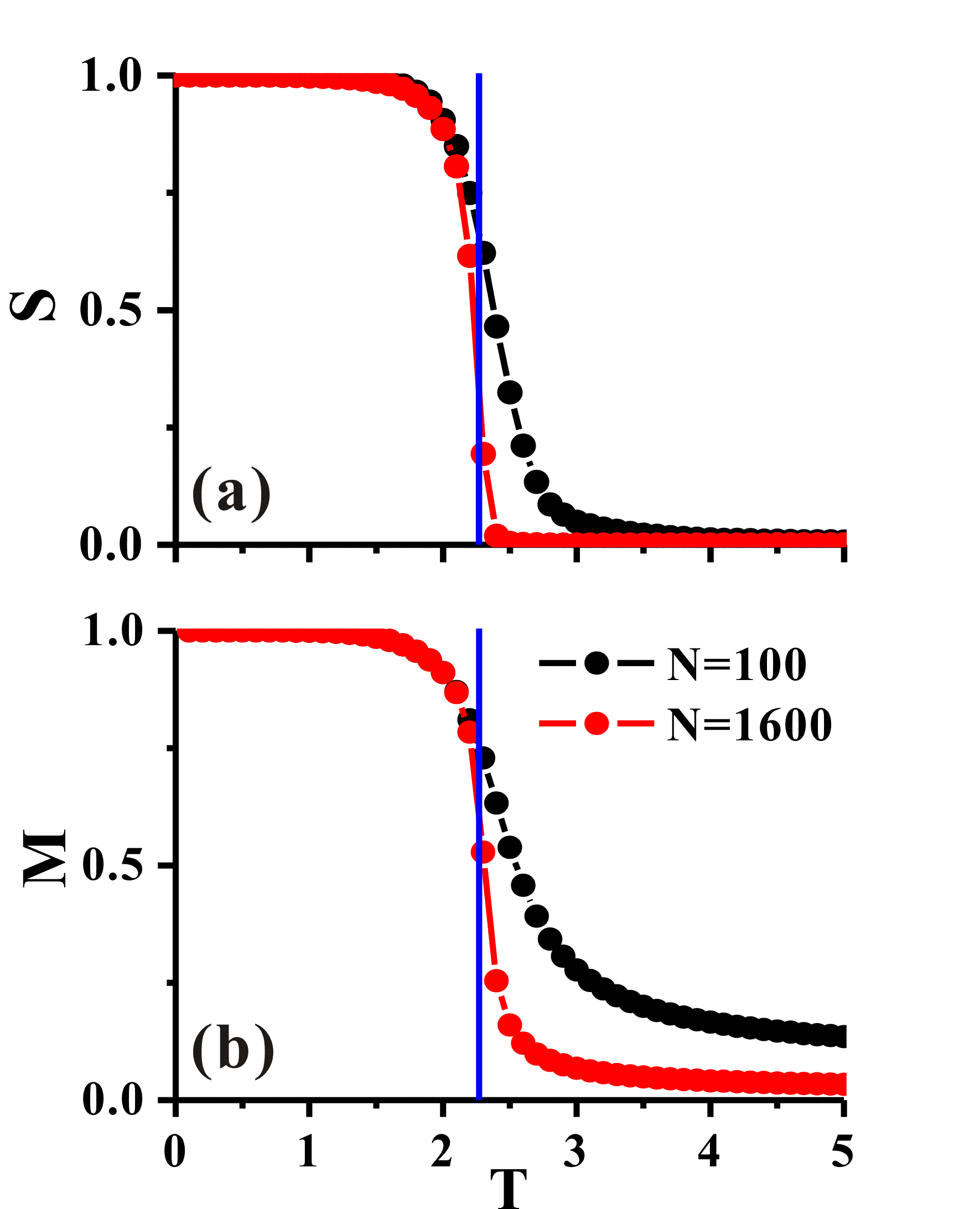}
\caption{ (a) Small-worldness $S$ and (b) Spontaneous magnetization
$M$ of 2D classical Ising model as a function of temperature $T$ for
the size of square lattice $N=100$ and $1600$. The blue vertical
line indicates the phase transition point $T_{c}$.}\label{fig3}
\end{figure}

This finding inspires us to carefully check whether or not the new
order parameter, small-worldness, is much easier to capture the
critical point of the classical Ising model on a square lattice.
Notice that here the definition of the adjacency matrix is replaced
by the spin-spin correlation function,
$\hat{A}_{ij}=\frac{|\langle\hat{\sigma}_{i}^{z}\hat{\sigma}_{j}^{z}\rangle|}{\max|\langle\hat{\sigma}_{i}^{z}\hat{\sigma}_{j}^{z}\rangle|}$.
In Fig.\ref{fig3}(a), one can see that given a lattice size the
small-worldness indeed needs much less effort to extract the
critical point than the spontaneous magnetization (compare to
Fig.\ref{fig3}(b)). This result can be understood as a consequence
for the bigger critical exponent of the small-worldness. In
practice, we thus present the efficiency of the small-worldness for
the numerical simulation of the order parameter in the 2D classical
Ising model.

\begin{figure}
\center
\includegraphics[height=2.4in,width=3.2in]{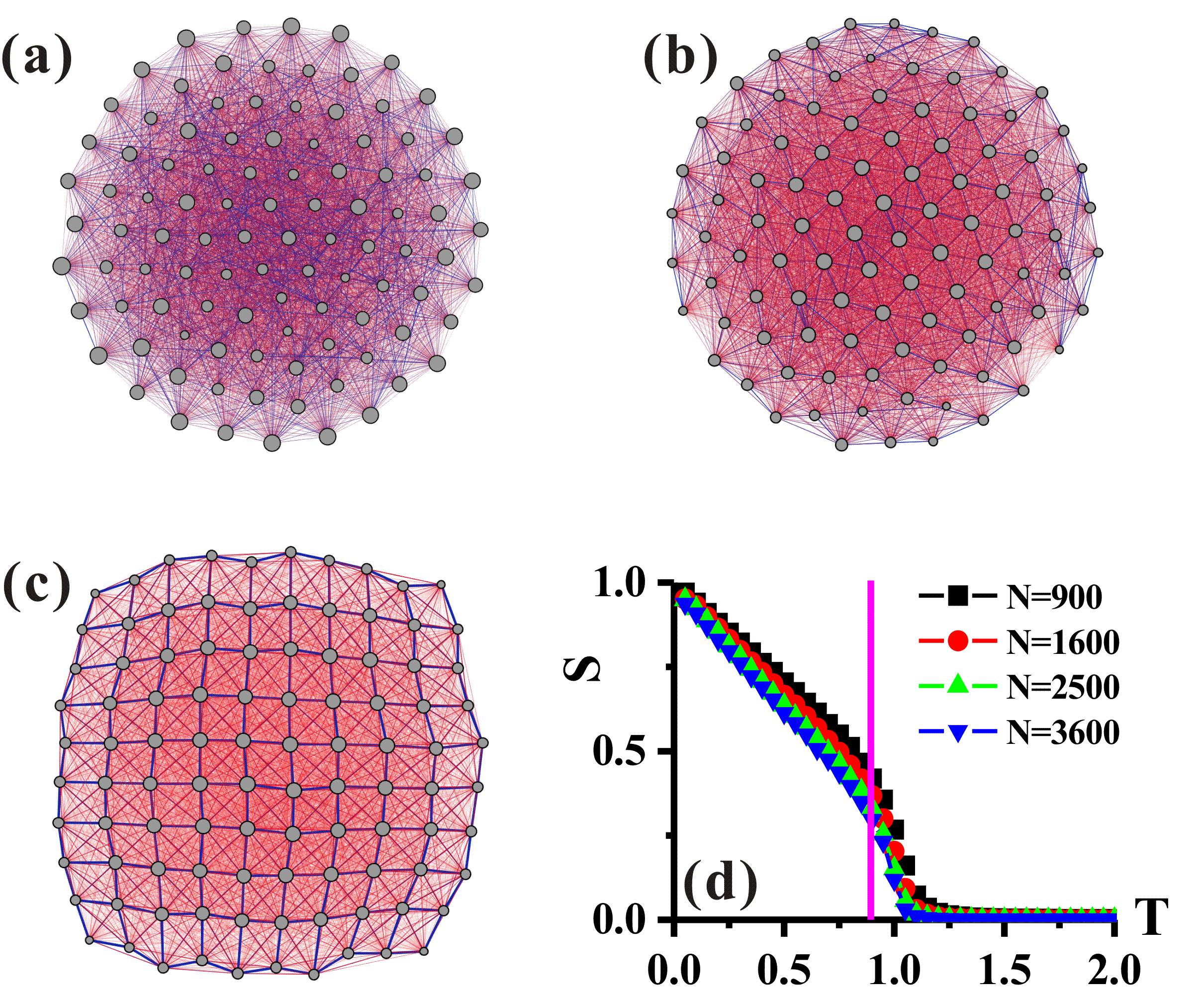}
\caption{Network representations of the classical rotor plane at
different temperature: (a) $T=0.1$, (b) $T=0.9$ and (c) $T=2.0$. The
lattice size $N=400$. The thickness of links represents the
amplitude of $C_{ij}$. The size of nodes stands for the node
strength. Here only $100$ nodes are plotted because of periodic
boundary condition. (d) Small-worldness $S$ of the classical rotor
plane as a function of temperature $T$ for different lattice size
$N$. The pink vertical line indicates the critical temperature
$T_{c}$ estimated by spin stiffness
\cite{HsieharXiv13}.}\label{fig4}
\end{figure}

\section{Rotor plane}\label{2dc}
The other example is the classical rotor plane, sometimes called 2D
classical XY model, in a square lattice of size $N$ described by
\cite{KosterlitzJPC73,KosterlitzJPC74}
\begin{eqnarray}
H_{XY}=-\sum_{\langle
i,j\rangle}\vec{S}_{i}\cdot\vec{S}_{j}=-\sum_{\langle
i,j\rangle}\cos\left(\theta_{i}-\theta_{j}\right),\label{e:equ1}
\end{eqnarray}
where $\theta_{i}$ is the angle of the 2D spin vector $\vec{S}_{i}$
at site $i$. Conventional long-range order, like ferromagnetism or a
crystal, is common in three dimensional systems. However, in 2D
systems with continuous symmetry as the model introduced above, the
true long-range order is completely washed out by thermal or quantum
fluctuations and only its topology remains.

In fact, the low-temperature phase forms a quasi-long-range order
originating from the power-law correlation decay. There is a phase
transition from this phase to the high-temperature disordered phase
whose correlations decay exponentially with distances. Such a
transition is known as the KT transition associated with the
disappearance of the quasi-long-range order. Near the topological
phase transition the system begins to lose spin stiffness that shows
up a universal jump near the KT transition temperature $T_{c}$. This
approach has been often used to numerically extract $T_{c}$
\cite{NelsonPRL77,WeberPRB87,HasenbuschJSM08,HsieharXiv13}. In the
following, a straightforward definition for the adjacency matrix is
the spin-spin correlation function
$C_{ij}\equiv\langle\cos\left(\theta_{i}-\theta_{j}\right)\rangle$
that can be calculated by using the standard Monte Carlo simulation.
We will show that the network topology also enables the
small-worldness to illustrate the KT phase transition.

\begin{figure}
\center
\includegraphics[height=3in,width=2.6in]{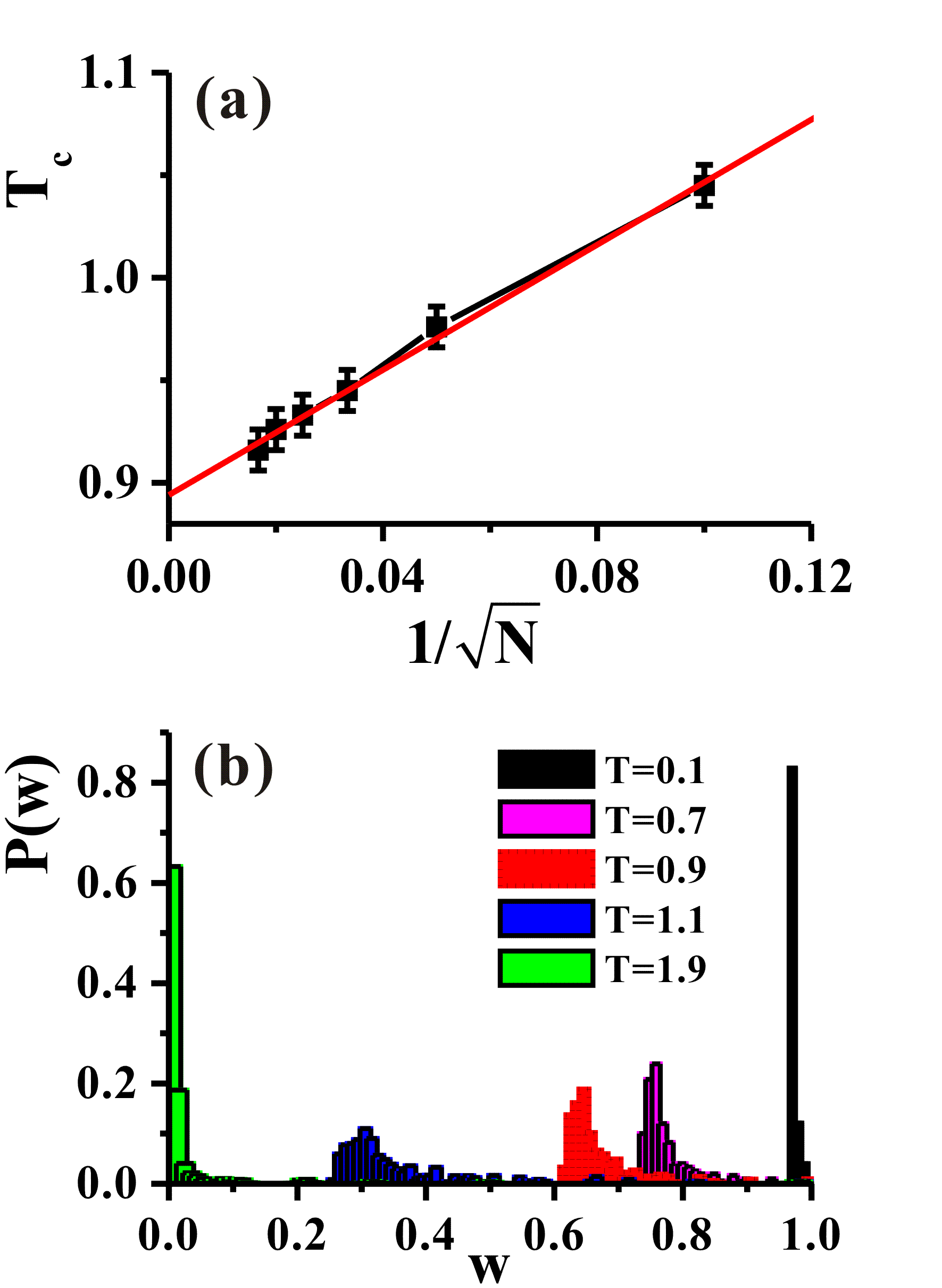}
\caption{(a) The finite size scaling of the critical temperature
$T_{c}$ determined from the deviation of the linear $T$ dependence
of $S$. (b) The probability distribution $P(w)$ of the link weight
$w$ for different $T$.}\label{fig5}
\end{figure}

Figure \ref{fig4}(a)-(c) show the temperature evolution of network
topologies in the rotor plane. As mentioned above, the
low-temperature topological phase enters the high-temperature
disordered phase at the KT transition temperature
($T_{c}\simeq0.9$). The low-temperature topology gives rise to the
power-law correlation decay, and hence links and nodes are arranged
in an irregular network pattern (see the example of $T=0.1$ as
illustrated in Fig.\ref{fig4}(a)). When temperature approaches the
KT transition point, some nodes begin to cluster and a
"crystal-like" structure is visualized as shown in
Fig.\ref{fig4}(b). The reason behind the structure is that near
$T_{c}$ the short-range correlations become more dominant than the
quasi-long-range order at low temperatures. At much higher
temperature ($T=2.0$), a square-lattice structure composed of the
strongest links appears in Fig.\ref{fig4}(c). Extremely short-range
correlations of the high-temperature disordered phase indicate much
larger link weights between the nearest-neighboring nodes so that
the network structure resembles a square lattice.

Spin stiffness has usually been used to determine the KT transition
temperature \cite{WeberPRB87}. Here we demonstrate that the
small-worldness $S$ defined in Eq.(\ref{e:equ5}) also have an
ability to detect the KT transition. In Fig.\ref{fig4}(d), we show
the critical behavior of the small-worldness vs temperature. One can
see that the temperature dependence of the small-worldness has a
linear decrease at low temperatures and an apparent deviation from
linearity around the KT transition. We thus define the critical
temperature $T_{c}$ where the small-worldness begins to deviate from
its linear behavior at low temperatures. In Fig.\ref{fig5}(a), we
estimate $T_{c}$($\simeq0.8940$) by finite size scaling, which is
very close to the value ($T_{c}\simeq0.8935$) adapted from the
universal jump of the spin stiffness \cite{HsieharXiv13}. This
agreement convinces us that a similar "jump" can be defined by using
the small-worldness as well. Moreover, similar to the Kitaev chain,
this result also implies that the small-worldness should be related
to the physical quantity corresponding to the second derivative of
the free energy with respect to a twist. As a result, we suggest
that the small-worldness can be also considered as a useful quantity
to characterize the phase transition, instead of spin stiffness.

Let us now discuss the weight distribution of network links in the
classical rotor plane. Unlike the Kitaev chain, Fig.\ref{fig5}(b)
shows that at low temperature ($T=0.1$) the weight distribution is
no longer similar to delta function but more like the log-normal
distribution. Most links center their weights around $w\sim1$, and
thus display the homogeneous distribution. As further increasing
temperature, the mean and variance of the distributions in
Fig.\ref{fig5}(b) becomes smaller and larger, respectively. It is
noteworthy that the heterogeneity of the weight distributions
appears before the KT transition. At high temperature, the peak of
the distribution is moved to $w\sim0$ with a long tail so that the
weight distributions become much more heterogeneous. As compared to
the topological phase transition in the Kitaev chain, the rotor
plane shows much broader weight distribution of network links for
all temperatures in the network space. It turns out that it is
rather difficult to come up with an order parameter in the classical
rotor plane. The same reasoning from the network topology can be
applied to other many-body systems without local order parameters.

\section{Conclusion and outlook}\label{conc}
In the Kitaev chain, we have illustrated that the small-worldness
clearly distinguish the topological non-trivial phase
($\mu<\mu_{c}$) and the topological trivial phase ($\mu>\mu_{c}$) in
the absence of symmetry distinction. In addition to the Chern number
\cite{ThoulessPRL82} and the Majorana number \cite{KitaevPU01}, we
have formulated another relevant topological quantity,
small-worldness, extracted from the network space allowing one to
characterize the topological phases. As compared to the critical
phenomena of the 2D classical Ising model, the critical exponent of
the small-worldness near $\mu_{c}$ is very close to the one of the
inverse of magnetic susceptibility, implying that the new order
parameter is within the universality Ising class and much easier to
capture the critical point than the spontaneous magnetization. In
addition, the transition of the weight distribution of network links
across the critical point can be also applied to other phase
transitions with the existence of conventional long-range orders.

In the classical rotor plane, we have found that the small-worldness
provides a shortcut to estimate the KT transition temperature.
According to our definition for the "universal jump" of the
small-worldness near the critical point, we have shown that the
transition temperature $T_{c}$ we obtained is almost the same as the
one estimated from the spin stiffness by previous large-scale Monte
Carlo simulations \cite{HsieharXiv13}. Again, this result supports
the finding in the Kitaev chain that the small-worldness should be
closely related to susceptibility or response to the external field.
We have also confirmed that the topological phase transition in the
2D classical rotor model can be characterized by using the weight
distribution of network links in the network representation.

In summary, we have proposed a novel complex network analysis for
computing the new topological invariant in 1D Kitaev model and
identifying the KT transition in 2D classical rotor model. By
defining the small-world network which have strongly coupled
small-world properties in these models, we have found that the
critical behavior of the many-body systems can be described by the
change of the weighted network topology. A network measure including
both local and non-local features, called small-worldness, has been
proven to be easier to investigate the quantum and classical phase
transitions. In particular, the picture behind the weight
distribution of network links provides significant information to
comprehend the generic phase transitions with/without local order
parameters in condensed matter systems. Based on the success of the
complex network analysis in this work, a very interesting direction
that we leave for the future is further detecting other
topologically ordered phases with the same topological entanglement
entropy in microscopic Hamiltonians, such as the toric code
\cite{KitaevAP03} and the double semion \cite{LevinPRB05} model.

\section*{Acknowledgments}
We thank C.-H. Lin, W.-F. Tsai and S.-M. Huang for helpful
discussions. Our special thanks go to S. P. Kou for bring novel
ideas to our attention. CPC is supported by Chinese Academy of
Engineering Physics and Ministry of Science and Technology. MCC is
supported by the National Science Council of Taiwan under the
contract number 102-2112-M-005-001-MY3.



\begin{thebibliography}{60}
\bibitem{smallworld}Note that in Ref.~\onlinecite{HumphriesPLos08} the small-worldness is normalized by its counterpart of a random network since they want to define a network which is a small world at the condition that $\langle s\rangle>1$. Here we use Eq.(\ref{e:equ5}) just to examine the relative small-worldness level between different phases.
\bibitem{gephi}The graph layout algorithm is implemented in Gephi.
\end{thebibliography}
\end{document}